# Acoustic amplification and bifurcation in a moving fluid

Zuwen Qian

Institute of Acoustics, Chinese Academy of Sciences, Beijing 100190, China

The quasi-accumulation solutions of acoustic wave in a moving fluid are obtained by using the Lagrange parameter variation method to solve the differential equation that describes the interaction between the acoustic waves and the flow. The results show that the nonlinear interaction causes the period-doubling followed by the odd multiple half-period bifurcation and all order subharmonics are generated subsequently, of which the amplitudes depend not only on the acoustic Mach number but also on the Mach number of the flow. The latter result indicates that the acoustic wave has been amplified by the momentum of the flow. The result also shows that the amplitudes of the generated subharmonics are proportional to the $m_1$ (the order number of the approximation) powers of the acoustic Reynolds number (and hence the Reynolds number of the flow). If the kinetic energy gained from momentum amplification is greater than the energy loss due to the acoustic attenuation, which means, the Reynolds number exceeds its critical value, a chain-reaction of the period-doubling followed by the odd multiple half-period bifurcation can continue to proceed so that the number of degrees of freedom in the flow increases infinitely resulting a chaos.

## 1. Introduction

According to the nonlinear sound propagation theory, a single frequency wave which propagates in a static medium will generate continuously all harmonic waves. However, due to the dissipative effects of the medium, the energy of the sound wave will eventually be attenuated out. In the previous publication [1], the author studied the nonlinear sound propagation in ideal media and obtained the so-called accumulation solutions. The results showed that the sound wave propagating in the flowing medium produces period-doubling bifurcations, of which the amplitude of each order of the subharmonics increases with the distance. During the process of the period-doubling bifurcation, the subharmonics continuously generate new subharmonics, in the form of a chain reaction. Since the medium concerned therein is ideal and the accumulation solutions obtained in reference [1] increase with increasing distance, thus, they do not convergent at infinity.

For actual media such as water, however, it has dissipation effect such as viscous dissipation,

the sound wave propagating in it will suffer an attenuation so that the divergence at infinity will no longer exist. In this paper, we introduce the viscous stress into the stress tensor and substitute it into the relevant equations in reference [1], and a further study for the bifurcation problem in a disturbed fluid is performed. The results show that the sound wave propagating in the moving fluid can be amplified in momentum, and the amplitude of the generated subharmonics depends on the Mach number of both the sound and the flow, especially, on the Reynolds number of the sound (or the flow). If the Reynolds number is large enough to counteract the effect of the sound attenuation due to the dissipation in media, it ensures that the period-doubling followed by the odd multiple half-period bifurcation for all subharmonics can continue to proceed in a chain reaction so that the number of degrees of freedom in the flow tends to infinity [2], and finally route to chaos.

When it comes to chaos, people think of turbulence. What are the criteria of the turbulence and its scenario to describe the processes of the laminar flow transit to the turbulence? In the field of mathematics, we can find the answer from [3-8] that is only logically reasonable, since the chaos is a necessary condition to form the turbulence. No publication is found providing proofs that the chaos is also a sufficient condition. In the field of physics, however, Landau proposed a criterion of turbulence [2], of which the solution contains an exponential term of time $e^{\gamma t}$ where $\gamma > 0$, even it does not converge in the time domain. Furthermore, the criterion also requires the number of freedom degrees of the motion must be infinite. On the other hand, we can find out the status of this problem from Refs. [9-11]. To summarize, the transition from laminar flow to turbulent flow is a very challenging problem [9], and it is also the sacred task of our researchers in physics.

## 2. Theory

As is well known, the one-dimensional form of stress tensor in viscous media should be written as

$$T_{ij} = T^{(l)} + T^{(s)}, \tag{1}$$

where

$$T^{(l)} = \rho v^2 \tag{2}$$

is the Reynolds stress, $T^{(s)}$ is the stress that can denoted by

$$T^{(s)} = p - C_0^2 \rho + (\mu_b + \frac{4}{3}\mu)\frac{\partial v}{\partial x}, \qquad (3)$$

where $p$ is the pressure, $\rho$ is the density, $C_0$ is the sound velocity, $\mu$ and $\mu_b$ are the shear viscosity coefficient and the volume viscosity coefficient, respectively. $v$ is the fluid particle velocity,

$$v = V_0 + v_0 \cos 2(\tau - \sigma). \qquad (4)$$

Substituting equations (1)-(4) into Lighthill equation yields

$$\frac{\partial^2 \rho}{\partial \tau^2} - \frac{\partial^2 \rho}{\partial \sigma^2} = \varepsilon \left\{ g_1(\sigma, \tau)\frac{\partial^2 \rho}{\partial \sigma^2} + g_2(\sigma, \tau)\frac{\partial \rho}{\partial \sigma} - g_3(\sigma, \tau)\rho \right\},$$

where

$$\varepsilon = Mm, \qquad (6)$$

and

$$\left.\begin{aligned}
g_1(\sigma, \tau) &= \frac{M}{m} + \frac{m}{2M} + 2\cos 2(\tau - \sigma) + \frac{m}{2M}\cos 4(\tau - \sigma) \\
g_2(\sigma, \tau) &= 8\sin 2(\tau - \sigma) + 4\frac{m}{M}\sin 4(\tau - \sigma) \\
g_3(\sigma, \tau) &= 8\left[\cos 2(\tau - \sigma) + \frac{m}{M}\cos 4(\tau - \sigma)\right]
\end{aligned}\right\}. \qquad (7)$$

Equation (5) describes the relationship of the interaction between sound waves and the flow. As we all know that the effect of stress $T^{(s)}$ on sound propagation is to add sound absorption to the amplitude of sound waves [6], that is, the amplitude of each order wave $A_{2n+1}^{(1)}$, $B_{2n+1}^{(1)}$ and $m$ should be rewritten into $A_{2n+1}^{(1)} e^{-(2n+1)\frac{\alpha}{k}\sigma}$, $B_{2n+1}^{(1)} e^{-(2n+1)\frac{\alpha}{k}\sigma}$ and $me^{-2\frac{\alpha}{k}\sigma}$ etc., respectively, where $\alpha$ is the sound absorption coefficient, $k$ is the wave number, and $\alpha \ll k$.

As did in Ref.[1], the solution of Eq.(5) can be written as

$$\rho = \rho^{(0)} + \varepsilon\rho^{(1)} + \varepsilon^2\rho^{(2)} + \ldots = \sum_{m_1}\varepsilon^{m_1}\rho^{(m_1)} \qquad (8)$$

and the zeroth order solution ($m_1 = 0$) can be

$$\rho^{(0)} = e^{-\frac{\alpha}{k}\sigma}\begin{cases} A_0\cos(\tau - \sigma) \\ B_0\sin(\tau - \sigma). \end{cases} \qquad (9)$$

In the following, we choose the cosine wave in equation (9) as the zeroth order solution (the case of sine wave can be treated similarly) and write it as $\rho^{(0c)}$. In the future, we use $\rho^{(m_1 c)}$ to denote the $m_1$ order approximate solution, where the superscript "c" means cosine wave excitation. When $m_1 = 1, 2, \ldots$, we write the corresponding density field as $\rho^{(1c)}$, $\rho^{(2c)}, \ldots$, which satisfy

$$\frac{\partial^2 \rho^{(m_1)}}{\partial \tau^2} - \frac{\partial^2 \rho^{(m_1)}}{\partial \sigma^2} = \left[ g_1(\sigma, \tau) \frac{\partial^2 \rho^{(m_1-1)}}{\partial \sigma^2} + g_2(\sigma, \tau) \frac{\partial \rho^{(m_1-1)}}{\partial \sigma} - g_3(\sigma, \tau) \rho^{(m_1-1)} \right]. \quad (10)$$

Substituting the cosine solution in equation (9) into (10) yields approximately

$$\frac{\partial^2 \rho^{(1c)}}{\partial \tau^2} - \frac{\partial^2 \rho^{(1c)}}{\partial \sigma^2} \approx -\sum_{n=0}^{2} A_{2n+1}^{(1)} e^{-(2n+1)\frac{\alpha}{k}\sigma} \cos[(2n+1)(\tau - \sigma)], \quad (11)$$

during the calculation, the approximation $\alpha \ll k$ was applied. In equation (11), $A_{2n+1}^{(1)}$ should be

$$\varepsilon A_1^{(1)} = 1^2 \left( M^2 + \frac{m^2}{2} + Mm \right) A_0, \quad \varepsilon A_3^{(1)} = 3^2 \left( Mm + \frac{m^2}{4} \right) A_0, \quad \varepsilon A_5^{(1)} = 5^2 \left( \frac{m^2}{4} A_0 \right). \quad (12)$$

By using the Lagrange parameter variation method, the solution of (11) are obtained as follows

$$\varepsilon \rho^{(1c)}(\tau, \sigma) = \varepsilon \sum_{n=0}^{2} A_{2n+1}^{(1)} e^{-(2n+1)\frac{\alpha}{k}\sigma} \frac{1}{4(2n+1)^2} \{2 \frac{k}{\alpha} \sin[(2n+1)(\tau - \sigma)] + \cos[(2n+1)(\tau - \sigma)]\}. (13)$$

Equation (13) shows us that a term proportional to $(k/\alpha)$ appears in the solution, of which its value is usually far much larger than 1, and we call it quasi-accumulation solution. By substituting this solution into equation (10), we can obtain the differential equation for $m_1 = 2$, which can denoted as

$$\frac{\partial^2 \rho^{(2c)}}{\partial \tau^2} - \frac{\partial^2 \rho^{(2c)}}{\partial \sigma^2} \approx \sum_{n=0}^{4} e^{-(2n+1)\frac{\alpha}{k}\sigma} \{\varepsilon^2 A_{2n+1}^{(2)} \cos(2n+1)(\tau - \sigma) + \varepsilon^2 B_{2n+1}^{(2)} \left( -\frac{k}{\alpha} \right) \sin(2n+1)(\tau - \sigma)\}. (14)$$

The solution of (14) can be

$$\varepsilon^2 \rho^{(2c)}(\sigma,\tau) = -\varepsilon^2 \sum_{n=0}^{4} \frac{1}{8(2n+1)^3} e^{-(2n+1)\frac{\alpha}{k}\sigma} \left\{ \begin{array}{l} \left( 2(2n+1) A_{2n+1}^{(2)} + \left[ 2(2n+1)^2 \left(\frac{k}{\alpha}\right)^2 - 1 \right] B_{2n+1}^{(2)} \right) \times \\ \cos(2n+1)(\sigma-\tau) + \\ + \left( 2(2n+1) B_{2n+1}^{(2)} - 4(2n+1)^2 A_{2n+1}^{(2)} \right) \left(-\frac{k}{\alpha}\right) \times \\ \sin(2n+1)(\sigma-\tau) \end{array} \right\}, \quad (15)$$

where the coefficients $A_{2n+1}^{(2)}$, $B_{2n+1}^{(2)}$ satisfy

$$\varepsilon^2 A_1^{(2)} = \frac{1}{4}\left\{ \left[3\left(M^2 + \frac{1}{2}m^2\right) - 5Mm\right]\varepsilon A_1^{(1)} + \left(\frac{11}{9}Mm - \frac{13}{36}m^2\right)\varepsilon A_3^{(1)} + \frac{19}{100}m^2\varepsilon A_5^{(1)} \right\}$$

$$\varepsilon^2 B_1^{(2)} = \frac{1}{2}\left\{ \left[\left(M^2 + \frac{1}{2}m^2\right) - Mm\right]\varepsilon A_1^{(1)} + \frac{1}{3}\left(Mm - \frac{1}{4}m^2\right)\varepsilon A_3^{(1)} + \frac{1}{20}m^2\varepsilon A_5^{(1)} \right\}$$

$$\varepsilon^2 A_3^{(2)} = \frac{3}{4}\left\{ \left(Mm - \frac{7}{4}m^2\right)\varepsilon A_1^{(1)} + \left(M^2 + \frac{1}{2}m^2\right)\varepsilon A_3^{(1)} + \frac{17}{25}Mm\varepsilon A_5^{(1)} \right\}$$

$$\varepsilon^2 B_3^{(2)} = \frac{3}{2}\left\{ 3\left(Mm - \frac{1}{4}m^2\right)\varepsilon A_1^{(1)} + \left(M^2 + \frac{1}{2}m^2\right)\varepsilon A_3^{(1)} + \frac{3}{5}Mm\varepsilon A_5^{(1)} \right\}$$

$$\varepsilon^2 A_5^{(2)} = -\frac{5}{16}m^2\varepsilon A_1^{(1)} + \frac{35}{36}Mm\varepsilon A_3^{(1)} + \frac{3}{4}\left(M^2 + \frac{1}{2}m^2\right)\varepsilon A_5^{(1)}$$

$$\varepsilon^2 B_5^{(2)} = \frac{5}{2}\left\{ \frac{5}{4}m^2\varepsilon A_1^{(1)} + \frac{5}{3}Mm\varepsilon A_3^{(1)} + \left(M^2 + \frac{1}{2}m^2\right)\varepsilon A_5^{(1)} \right\} \quad (16)$$

$$\varepsilon^2 A_7^{(2)} = \frac{7}{4}\left\{ \frac{5}{36}m^2\varepsilon A_3^{(1)} + \frac{13}{25}Mm\varepsilon A_5^{(1)} \right\}, \quad B_7^{(1)} = \frac{49}{2}\left\{ \frac{1}{12}m^2\varepsilon A_3^{(1)} + \frac{1}{5}Mm\varepsilon A_5^{(1)} \right\}$$

$$\varepsilon^2 A_9^{(2)} = \frac{99}{400}m^2\varepsilon A_5^{(1)}, \quad B_9^{(2)} = \frac{81}{40}m^2\varepsilon A_5^{(1)}$$

Similar to the first-order solution (10), the term of quasi-accumulation solution $(k/\alpha)^2$ appears in formula (15). In addition, the equation (16) shows us that $A_{2n+1}^{(1)}$, $A_{2n+1}^{(2)}$, $B_{2n+1}^{(1)}$ and $B_{2n+1}^{(2)}$ depend on both the Mach number of sound and the Mach number of flow.

### 3. Discussion

**I. Period-doubling followed by the odd multiple half-period bifurcation**

From the results abovementioned, we can also see that the sound propagation in this type of medium is a period-doubling followed by the odd multiple **half-period** bifurcation process. The original excitation wave is $v_0 e^{-2\frac{k}{\alpha}\sigma} \cos 2(\tau-\sigma)$ and the subharmonics are the combination of $e^{-(2n+1)\frac{\alpha}{k}\sigma}\sin[(2n+1)(\tau-\sigma)]$ and $e^{-(2n+1)\frac{\alpha}{k}\sigma}\cos[(2n+1)(\tau-\sigma)]$, $n=0,1,2,...$ . And following figure describes the bifurcation process:

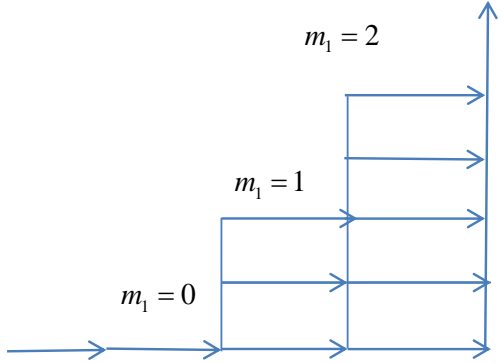

**Figure 1. bifurcation**

where $m_1$ is the number of the order of the approximate solutions. In the figure, the first arrow represents the original excitation wave, the second arrow represents the generated first sub harmonic ($m_1=0$), and its phase factor is $(\tau-\sigma)$; when $m_1=1$, there are three arrows to indicate the three generated subharmonics, of which the phase factors are $(\tau-\sigma), 3(\tau-\sigma)$ and $5(\tau-\sigma)$, respectively; when $m_1=2$, there are five arrows to indicate the five generated subharmonics, of which the phase factors are 1, 3, 5, 7 and 9 times of the $(\tau-\sigma)$, respectively. From the bifurcation process, we can see that the bifurcation described by the first arrow is period doubling, and the next bifurcations are different from the period doubling bifurcation, thus, we call the latter the odd multiple half-period bifurcation. The highest growth term in their quasi-cumulative solution is $(k/\alpha)^{m_1}$. With the increase of Reynolds number, the amplitudes of all subharmonics increase and the number of subharmonics (degrees of freedom) increases infinitely.

**II. "Flow" amplifies "sound"**

By (13) and (15) it can be seen that due to the nonlinear interaction between the flow in the moving fluid and the sound waves of which the amplitudes only depend on their own Mach number before the interaction the amplitude of the generated harmonics depend not only on the acoustic Mach number, but also on the Mach number of the flow after the interaction. On the other hand, from the solutions of equations (9), (13) and (15), it can also be seen that the subharmonics generated by the nonlinear interaction of sound waves consists of two parts, one of which is directly proportional to the term $(k/\alpha)$ or $(k/\alpha)^2$ that were called quasi-accumulation solutions. When the frequency of sound wave is 100 Hz, $k/\alpha$ (for water) can reach the numerical range of $10^9$ (Nepers)$^{-1}$. Therefore, these terms in (13) and (15) are much larger than other terms. On the other hand, due to the existence of sound absorption in the medium, the other terms of the solutions in (13) and (15) will be attenuated and eventually extinguished, leaving only the quasi-cumulative solution terms that proportional to $k/\alpha$ or $(k/\alpha)^2$. Thus, (13) and (15) can be rewritten as

$$\varepsilon\rho^{(1c)}(\tau,\sigma) = \varepsilon\sum_{n=0}^{2} A_{2n+1}^{(1)} e^{-(2n+1)\frac{\alpha}{k}\sigma} \frac{1}{2(2n+1)^2} \frac{k}{\alpha} \sin(2n+1)(\tau-\sigma) \quad (17)$$

and

$$\varepsilon^2\rho^{(2c)}(\sigma,\tau) = -\varepsilon^2\sum_{n=0}^{4} \frac{1}{4(2n+1)^2} e^{-(2n+1)\frac{\alpha}{k}\sigma} \left\{ \begin{array}{l} \left[(2n+1)\left(\frac{k}{\alpha}\right)^2\right] B_{2n+1}^{(2)} \cos(2n+1)(\sigma-\tau) + \\ + \left( \begin{array}{l} 2(2n+1)A_{2n+1}^{(2)} - \\ -B_{2n+1}^{(2)} \end{array} \right)\left(\frac{k}{\alpha}\right) \sin(2n+1)(\sigma-\tau) \end{array} \right\}. \quad (18)$$

It is easy to prove that the quantity $k/\alpha = \frac{2}{m}Ra$, where $Ra$ is the acoustic Reynolds number of the medium. Since $A_{2n+1}^{(2)}$ and $B_{2n+1}^{(2)}$ depend on the Mach numbers $M$ and $m$, indicating that, $A_{2n+1}^{(1)}\frac{k}{\alpha} \propto \text{Re}$, $B_{2n+1}^{(2)}\left(\frac{k}{\alpha}\right)^2 \propto (\text{Re})^2$, where Re is the Reynolds number of the moving fluid. It can also be seen that the interaction between sound and flow leads to the acoustic wave gain momentum amplification from the moving fluid. On the other hand, the generated subharmonics will be attenuated due to the dissipation of the medium. If the energy obtained exceeds the attenuated energy, the sound wave will continue to propagate and bifurcate continuously. Moreover, with the increase of Reynolds number, the bifurcation process is enhanced. On the contrary, if the energy absorbed is greater than the energy obtained, the propagation and bifurcation of the sound wave will gradually weaken and eventually extinguish.

**Amplification gain**

It can be seen from equations (9) and (18) that for $m_1 = 2$ the amplification gain D of the flow to the first subharmonic is

$$D = \frac{1}{64}\left(\frac{k}{\alpha}\right)^2 \left(8M^4 + 24M^2 m^2 + 3m^4\right) \quad (19)$$

When m = 0 and M = m, it can be calculated from equation (19) that the amplitude of the first subharmonic in the quasi accumulation solution is amplified by nearly 12 times due to the appearance of flow. Reviewing the experimental results of reference [1], the first subharmonic can be only observed in disturbed water, but not in stationary water. This paper gives it a powerful explanation.

As is well known that the first subharmonic can be amplified by the current only when the gain $D$ must be greater than 1. For water, when the frequency ranges from $f = 10^2$ Hz to $10^6$ Hz, the flow velocity $V_0$ should be greater than 1.95m/s and 0.06m/s, respectively, in case of $M \gg m$.

An example will be given as follows. When m = 0 and M = m, we calculated separately the amplitude of the first subharmonic in the quasi-accumulation solution (18), the result indicates that the amplitude of the first subharmonic is amplified by nearly 12 times due to the appearance of a flow. Let's take a look at the figures 1 and 2 in Reference [1](cf. Attached figures), which show that the first subharmonic can only be observed in disturbed water, but not in stationary water. This provides a convincing explanation to these experimental results.

## 4. Conclusions

In this paper, the author applied the Lagrange parameter variation method to solve the differential equation that describes the interaction between sound waves and the flow and the quasi-accumulation solutions which is proportional to $m_1$ power of the Reynolds number are obtained, where $m_1$ is the number of the order of the approximate solutions. The results show that the wave propagating in the moving fluid produces the period-doubling followed by the odd multiple half-period bifurcation of the subharmonics, and the Mach number of the moving fluid generates momentum amplification to the sound wave, and in the same time, each subharmonic is also subject to the attenuation of the medium. Therefore, Reynolds number is a physical quantity to measure the competition between Mach momentum amplification and sound attenuation. If the kinetic energy obtained from momentum amplification is greater than the energy attenuated, the process of the bifurcation due to the nonlinear interaction between moving fluid and sound wave can proceed continuously. With the increase of Reynolds number, each subharmonic itself will generate new subharmonics during its propagation process and a chain reaction bifurcation is formed, which can make the system transition to chaos due to the infinite increase of the number of degrees of freedom in the flow.

[11] Friedrich H. Busse**,** Visualizing the Dynamics of the Onset of Turbulence, SCIENCE 2004; **305**: 1574-1575.

[12] Kinsler L E, Frey A R, Coppens A B and Sanders J V, Fundamentals of Acoustics: Absorption and attenuation of sound Ch. 8, 4[th] Edition, New York, John Wiley & Sons, Inc.; 2000.

■■■■■■■■■■■■■■■■■■■■■■■■■■■■■■■■■■■■■■■■■■■■■■■■■■■■■■■■■■■■■■■■

Attached figures

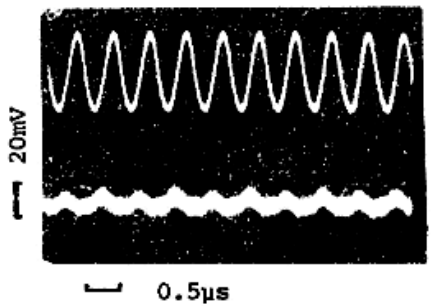 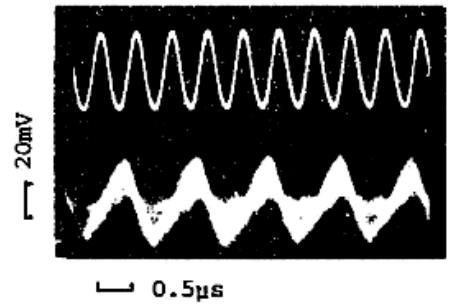

Figure 1. The signals in static medium:          Figure 2. The signals in disturbed medium
Fundamental frequency signal (upper)            Fundamental frequency signal (upper)
First subharmonic signal   (lower)              First subharmonic signal   (lower)
(From figure 2 of Ref.[1])                      (From figure 3 in Ref.[1])